%% file: main.tex
\def\arxiv{}
\newif\iflinenums
\linenumstrue
\def\docopts{manuscript}
\def\docclass{aastex}
\def\figwidth{0.5\textwidth}

\ifdefined\arxiv
  \linenumsfalse
  \def\docopts{aps,prl,reprint,superscriptaddress,floatfix,nofootinbib,showkeys,showpacs}
  \def\docclass{revtex4-1}
  \def\figwidth{\columnwidth}
  
  \def\tblnotemark{\footnotemark}
  \def\tblnotetext{\footnotetext}
\fi

\ifdefined\aastex
  \linenumsfalse
  \def\docopts{preprint}
  \def\docclass{aastex}
  \def\figwidth{0.7\textwidth}
  
  \def\tblnotemark[#1]{\tablenotemark{#1}}
  \def\tblnotetext[#1]{\tablenotetext{#1}}
\fi

\ifdefined\apj
  \linenumsfalse
  \def\docopts{iop,revtex4-1}
  \def\docclass{hackemulateapj}
  \def\figwidth{\columnwidth}
  
  \def\tblnotemark[#1]{\tablenotemark{#1}}
  \def\tblnotetext[#1]{\tablenotetext{#1}}
\fi

\ifdefined\lat
  \linenumstrue
\fi

\documentclass[\docopts]{\docclass}

\pdfoutput=1

\usepackage{aas_macros}
\usepackage{tabularx}
\usepackage{xspace}
\usepackage{listings}
\usepackage{lineno}
\iflinenums
  \linenumbers
\fi
\usepackage[dvipsnames,svgnames]{xcolor} 
\usepackage{graphicx}
 
\usepackage{natbib}
\usepackage{hyperref}
\usepackage{amsmath}
\usepackage{multirow}
\hypersetup{    
  colorlinks      = true,
  linkcolor       = {black},
  linkbordercolor = {white},
  citecolor       = {black},
  citebordercolor = {white},
  urlcolor        = {blue},
  urlbordercolor  = {white},
}

\input{commands.tex}

\graphicspath{{./}{../figures/}}

\begin{document}

\title{Search for Gamma-Ray Emission from DES Dwarf Spheroidal Galaxy Candidates with Fermi-LAT Data}
\input{\authlist}

\begin{abstract}
Due to their proximity, high dark-matter content, and apparent absence of non-thermal processes, Milky Way dwarf spheroidal satellite galaxies (dSphs) are excellent targets for the indirect detection of dark matter. 
Recently, \nobjs new dSph candidates were discovered using the first year of data from the Dark Energy Survey (DES).
We searched for gamma-ray emission coincident with the positions of these new objects in six years of \Fermi Large Area Telescope data. 
We found no significant excesses of gamma-ray emission.
Under the assumption that the DES candidates are dSphs with dark matter halo properties similar to the known dSphs, we computed individual and combined limits on the velocity-averaged dark matter annihilation cross section for these new targets.
If the estimated dark-matter content of these dSph candidates is confirmed, they will constrain the annihilation cross section to lie below the thermal relic cross section for dark matter particles with masses $\lesssim 20 \GeV$ annihilating via the \bbbar or \tautau channels.

\end{abstract}

\maketitle

\section{Introduction}\label{sec:intro}

In the standard model of cosmology, dark matter (DM) is the dominant component of matter in the universe.
Weakly interacting massive particles (WIMPs) are an attractive candidate to constitute some or all of DM \citep[\eg,][]{Bertone:2004pz,Feng:2010gw}. 
If WIMPs are in thermal equilibrium in the early universe and have a velocity-averaged annihilation cross section of $\sigmav \sim \relic$, their relic abundance can account for the observed DM abundance measured today~\citep[\eg,][]{Steigman:2012nb}. 
WIMPs may continue to annihilate in regions of high DM density to produce energetic Standard Model particles that can be detected as indirect signatures of DM.
These indirect searches complement terrestrial searches for DM in accelerator and direct detection experiments~\citep[\eg,][]{Bauer:2013ihz}. 

Gamma rays are one product of WIMP annihilations~\citep{Baltz:2008wd,Bringmann:2012ez}; they may be produced directly or in a shower of secondary particles.
Depending on the WIMP mass, these gamma rays could be detectable with the \Fermi Large Area Telescope (LAT)~\citep{Atwood:2009ez}. 

The integrated gamma-ray flux in a specific energy range ($E_{\min} < E < E_{\max}$) and region of interest (ROI) on the sky from DM annihilation is given by
\begin{equation}
\begin{aligned}
       \phi_s(\Delta\Omega) =
    & \underbrace{ \frac{1}{4\pi} \frac{\sigmav}{2m_{\DM}^{2}}\int^{E_{\max}}_{E_{\min}}\frac{\text{d}N_{\gamma}}{\text{d}E_{\gamma}}\text{d}E_{\gamma}}_{\rm particle~physics}\\
    &    \times
    \underbrace{\vphantom{\int_{E_{\min}}} \int_{\Delta\Omega}\int_{\rm l.o.s.}\rho_{\DM}^{2}(\vect{r})\text{d}s\text{d}\Omega }_{\rm \Jfactor}\,,
\end{aligned}
\label{eqn:annihilation}
\end{equation}
\noindent where the first term encompasses the particle properties of the DM, while the second term (the so-called ``\Jfactor'') incorporates information about the distribution of DM along the line of sight. 
Specifically, $\mDM$ is the DM particle mass, $\text{d}N_{\gamma} / \text{d}E_{\gamma}$ is the differential gamma-ray yield per annihilation summed over all final states, $\Delta\Omega$ is the solid angle of the ROI, and  $\rho_{\DM}(\vect{r})$ is the DM density.
 
Current N-body cosmological simulations of Milky Way-sized regions predict the existence of thousands of Galactic DM overdensities called subhalos~\citep{Diemand:2008in,Springel:2008cc}. 
Luminous Milky Way dwarf spheroidal satellite galaxies (dSphs) are believed to reside in a subset of the most massive subhalos.
The Milky Way dSphs are especially promising targets for indirect DM searches due to their large dark matter content, low diffuse Galactic $\gamma$-ray foregrounds, and lack of conventional astrophysical $\gamma$-ray production mechanisms~\citep{McConnachie:2012vd}.
Several searches for gamma-ray emission from known dSphs have been performed using LAT data, none of which has resulted in a positive detection \citep[\eg,][]{Abdo:2010ex,Ackermann:2011wa,GeringerSameth:2011iw,Mazziotta:2012ux,Ackermann:2013yva,Geringer-Sameth:2014qqa,Ackermann:2015zua}.

The census of known Milky Way satellites is certainly incomplete. 
Prior to the Sloan Digital Sky Survey (SDSS) \citep{York:2000gk}, there were ten dSphs known to orbit the Milky Way (called classical dwarfs). 
The deep and systematic coverage of the northern celestial hemisphere by SDSS has more than doubled the number of known Milky Way satellites~\citep{McConnachie:2012vd}.
Additionally, SDSS data led to the discovery of a new population of ``ultra-faint'' satellite galaxies, which were found to be the most DM dominated objects known~\citep{2007ApJ...670..313S,2008ApJ...678..614S,2009ApJ...692.1464G}.
The Dark Energy Survey (DES)~\citep{Abbott:2005bi} is a southern-hemisphere optical survey expected to find new dSphs~\citep{Tollerud:2008ze,Hargis:2014kaa}, which would increase the sensitivity of searches for particle DM \citep{He:2015}.

Photometric survey data can be used to identify stellar overdensities associated with satellite dwarf galaxies or globular clusters. 
Satellite galaxies require DM to explain their observed kinematics, while the mass of globular clusters can be accounted for by their visible matter alone. 
Globular clusters can be distinguished from dwarf galaxies based on spectroscopic measurements~\citep{Willman:2012uj}.
The range of stellar  metallicities in globular clusters is narrower than that observed in dSph galaxies.
Though globular clusters and satellite galaxies may possess similar stellar velocity dispersions, the larger spatial extent of dwarf galaxies implies that they are DM-dominated.

The first internal annual release of DES data (Y1A1) covers $\roughly1{,}800 \deg^2$ in the southern hemisphere ($\roughly1{,}600 \deg^2$ not overlapping with SDSS).\footnote{\url{http://data.darkenergysurvey.org/aux/releasenotes/DESDMrelease.html}}
Recent studies of the Y1A1 data set have revealed \nobjs new dSph candidates \citep{Bechtol:2015wya,Koposov:2015cua}.\footnote{\citet{Koposov:2015cua} find a ninth candidate inside the DES year-one imaging footprint but outside the Y1A1 coadd catalog.}
Since the LAT continuously surveys the entire sky, LAT data collected over the duration of the mission can be used to search for gamma-ray emission from the DES dSph candidates.

\section{Discovery of New dSph Candidates with DES}\label{sec:DES}
 
Current and near-future deep wide-field optical imaging surveys have the potential to discover many new ultra-faint Milky Way satellites~\citep{Tollerud:2008ze,Hargis:2014kaa,He:2015}.
The ensemble of PanSTARRS~\citep{Kaiser:2002zz}, the SkyMapper Southern Sky Survey~\citep{Keller:2007cd}, DES~\citep{Abbott:2005bi}, and the Large Synoptic Survey Telescope~\citep{Ivezic:2008fe} will explore large areas of the sky to unprecedented depths.
Here, we focus on a set of dSph candidates recently found in first-year DES data.

Details regarding the first-year DES data set and techniques to search for ultra-faint dSphs are provided in~\citet{Bechtol:2015wya} and~\citet{Koposov:2015cua}.
Briefly, a dSph candidate is identified as a statistically significant arcminute-scale overdensity of resolved stars consistent with an old (${>}\,10\Gyr$) and metal-poor ($Z \sim 0.0002$) stellar population. 
A variety of search techniques have been applied to the first-year DES data, including visual inspection of DES images, thresholding stellar density maps, scanning with optimized spatial filters, and automated matched-filter maximum-likelihood algorithms.
The physical characteristics of dSph candidates (\eg, centroid position, distance, and spatial extension) can be inferred by fitting the spatial and color-magnitude distributions of the stars.
\tabref{DESdwarfs} provides a summary of the \nobjs dSph candidates reported by~\citet{Bechtol:2015wya}.

\begin{table}[ht]
  \caption{DES dSph Candidates and Estimated J-factors}
  \begin{center}
  \begin{tabular}{lccc}
\hline\hline
\label{tab:DESdwarfs}
Name & ($\ell$, $b$)\tblnotemark[1] & Distance\tblnotemark[2]  &  $\log_{10}(\text{Est.J})$\tblnotemark[3] \\[3pt]  
     &    $\deg$     & $\kpc$   &  $\log_{10}(\frac{\GeV^2}{\cm^{5}})$ \\[4pt]
\hline
\DESJf & $(275.0,-59.6)$ & 95  & 18.3 \\
\DESJc & $(271.4,-54.7)$ & 87  & 18.4 \\
\DESJa & $(266.3,-49.7)$ & 32  & 19.3 \\
\DESJb & $(249.8,-51.6)$ & 330 & 17.3 \\
\DESJd & $(257.3,-40.6)$ & 126 & 18.1 \\
\DESJh & $(347.2,-42.1)$ & 69  & 18.6 \\
\DESJe & $(328.0,-52.4)$ & 58  & 18.8 \\
\DESJg & $(323.7,-59.7)$ & 95  & ~~~18.3

\tblnotetext[1]{\,Galactic longitude and latitude.}
\tblnotetext[2]{\,We note that typical uncertainties on the distances of dSphs are 10--15\%.}
\tblnotetext[3]{\,\Jfactors are calculated over a solid angle of $\Delta\Omega \sim 2.4 \times 10^{-4} \sr$ (angular radius $0\fdg5$). See text for more details.}
\\\hline\hline
\end{tabular}
\end{center}
\end{table}

\section{LAT Analysis}\label{sec:LAT}

\begin{figure}[t]
  \begin{centering}
  \includegraphics[width=\figwidth]{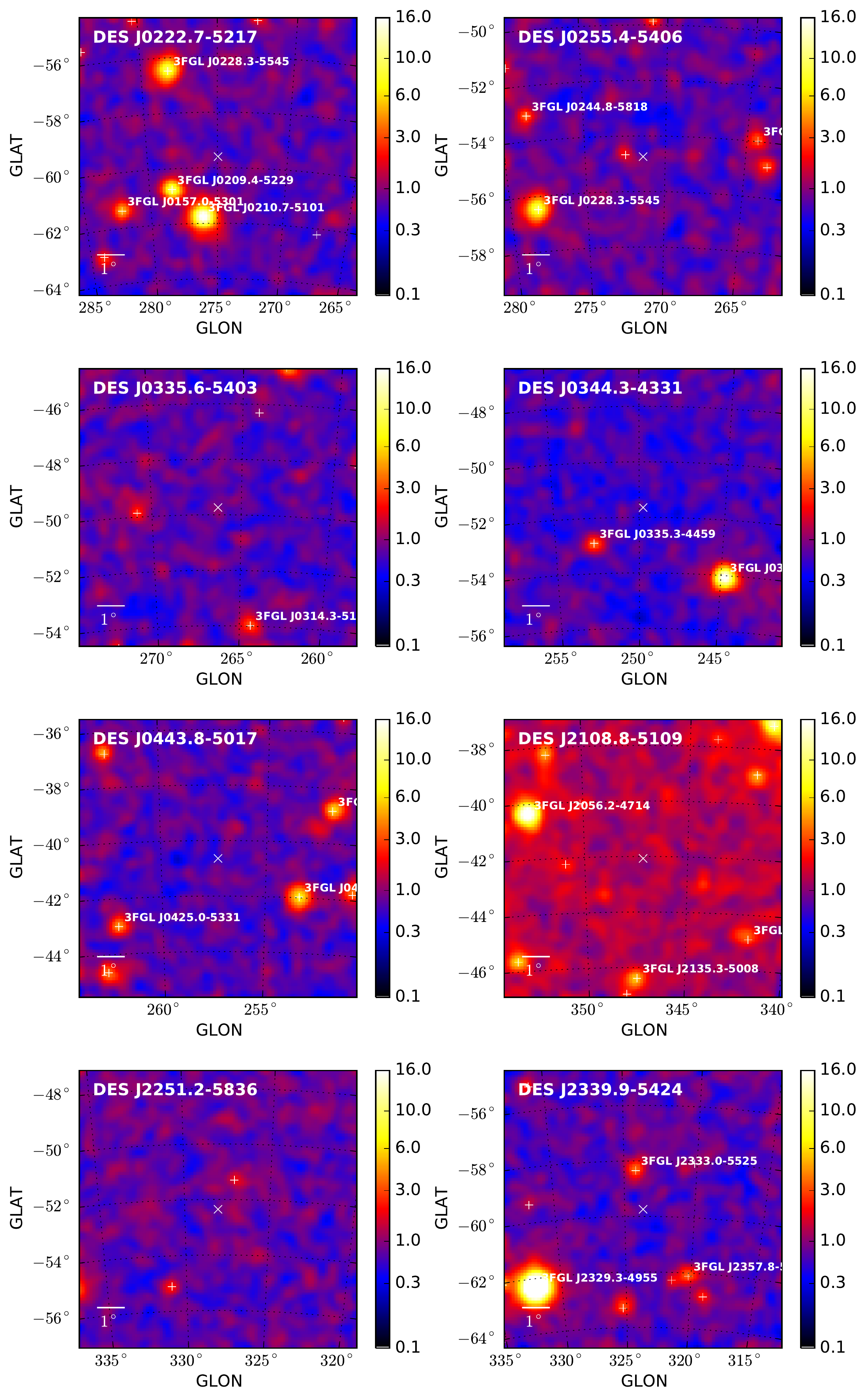}
  \caption{\label{fig:LATCounts} LAT counts maps in $10^{\circ}\times10^{\circ}$ ROI centered at each DES dSph candidate (white ``$\times$'' symbols), for $E > 1 \GeV$, smoothed with a $0\fdg25$ Gaussian kernel. All 3FGL sources in the ROI are indicated with white ``$+$'' symbols, and those with $\TS > 100$ are explicitly labeled.}

\end{centering}
\end{figure}

To search for gamma-ray emission from these new dSph candidates, we used six years of LAT data (2008 August 4 to 2014 August 5) passing the \irf{P8R2\_SOURCE} event class selections from $500\MeV$ to $500\GeV$.
The low-energy bound of 500 \MeV is selected to mitigate the impact of leakage from the bright limb of the Earth because the point spread function broadens considerably below that energy.
The high-energy bound of 500 \GeV is chosen to mitigate the effect of the increasing residual charged-particle background at higher energies~\citep{Ackermann:2014usa}.
Compared to the previous iteration of the LAT event-level analysis, \irf{Pass 8} provides significant improvements in all areas of LAT analysis; specifically the differential point-source sensitivity improves by 30\%--50\% in \irf{P8R2\_SOURCE\_V6} relative to \irf{P7REP\_SOURCE\_V15} \citep{Atwood:2013rka}. 
To remove gamma rays produced by cosmic-ray interactions in the Earth's limb, we rejected events with zenith angles greater than $100^{\circ}$.  
Additionally, events from time intervals around bright gamma-ray bursts and solar flares were removed using the same method as in the 4-year catalog analysis (3FGL)~\citep{Ackermann:2015hja}. 
To analyze the dSph candidates in \tabref{DESdwarfs}, we used $10^{\circ}\times 10^{\circ}$ ROIs centered on each object. 
Data reduction was performed using \stools\ version 10-01-00.\footnote{\url{http://fermi.gsfc.nasa.gov/ssc/data/analysis/software/}}
\figref{LATCounts} shows smoothed counts maps around each candidate for energies ${>}\,1\GeV$.
The candidate dSphs reside in regions of the sky where the diffuse background is relatively smooth.
With the exception of \DESJc, all the objects are located more than 1\degree from 3FGL background sources (\DESJc is located 0\fdg63 from 3FGL\,J0253.1$-$5438).

We applied the search procedure presented in \citet{Ackermann:2015zua} to the new DES dSph candidates. 
Specifically, we performed a binned maximum-likelihood analysis in 24 logarithmically spaced energy bins and $0\fdg1$ spatial pixels. 
Data are additionally partitioned in one of four PSF event types, which are combined in a joint-likelihood function when performing the fit to each ROI \citep{Ackermann:2015zua}.

\begin{figure*}
  \begin{center}
  \includegraphics[width=\textwidth]{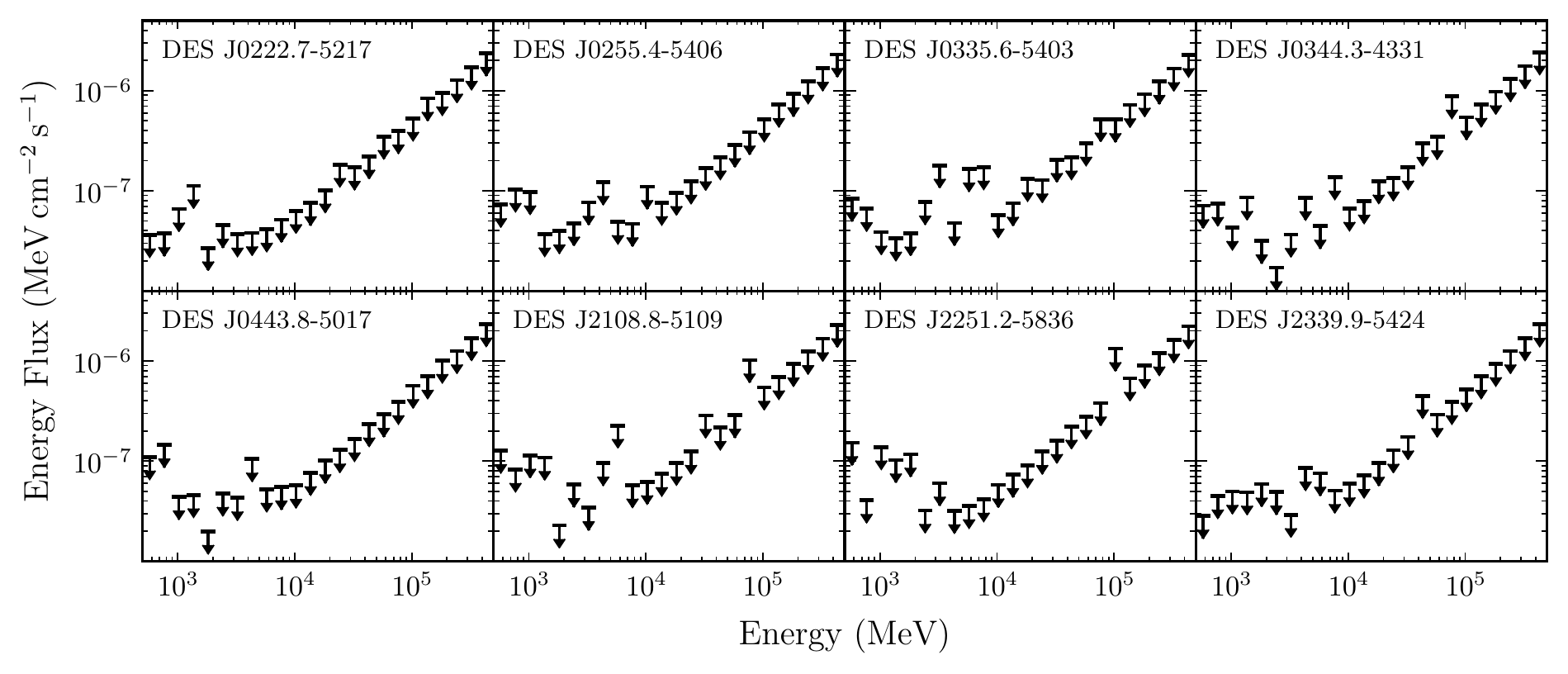}
  \caption{\label{fig:sed} Bin-by-bin integrated energy-flux upper limits at 95\% confidence level for the \nobjs DES dSph candidates modeled as point-like sources.}
\end{center}
\end{figure*}

We used a diffuse emission model based on the model for Galactic diffuse emission derived from an all-sky fit to the \irf{Pass 7 Reprocessed} data,\footnote{\url{http://fermi.gsfc.nasa.gov/ssc/data/access/lat/BackgroundModels.html}} but with a small ($<10\%$) energy-dependent correction to account for differences in the \irf{Pass 8} instrument response.\footnote{Standard LAT analyses treat the diffuse emission model as being defined in terms of true energy, but the model was necessarily derived from the measured energies of events. This implies a weak dependence of the model on the instrument response functions. The correction applied to the diffuse emission model accounts for the different energy dependence of the effective area and energy resolution between \irf{Pass 7 Reprocessed} and \irf{Pass 8}.}
Additionally, we model extragalactic gamma-ray emission and residual charged particle contamination with an isotropic model fit to the \irf{Pass 8} data.
These models will be included in the forthcoming public \irf{Pass 8} data release.
Point-like sources from the recent 3FGL catalog~\citep{Ackermann:2015hja} within $15^{\circ}$ of the ROI center were also included in the ROI model.
The spectral parameters of these sources were fixed at their 3FGL catalog values.
The flux normalizations of the Galactic diffuse and isotropic components and 3FGL catalog sources within the $10^{\circ}\times 10^{\circ}$ ROI were fit simultaneously in a binned likelihood analysis over the broadband energy range from 500\MeV to 500\GeV. The fluxes and normalizations of the background sources are insensitive to the inclusion of a putative power-law source at the locations of the DES dSph candidates, as expected when there is no bright point source at the center of the ROI.

In contrast to \citet{Ackermann:2015zua}, we modeled the dSph candidates as point-like sources rather than spatially extended Navarro, Frenk and White (NFW) DM density profiles~\citep{Navarro:1996gj}. 
This choice was motivated by the current uncertainty in the spatial extension of the DM halos of these new objects.
Previous studies have shown that the LAT flux limits are fairly insensitive to modeling dSph targets as point-like versus spatially extended sources~\citep{Ackermann:2013yva}.
Following the procedure of \citet{Ackermann:2015zua}, we fit for excess gamma-ray emission associated with each target in each energy bin separately to derive flux constraints that are independent of the choice of spectral model. 
Within each bin, we model the putative dSph source with a power-law spectral model ($dN/dE$ $\propto E^{-\Gamma}$) with spectral index of $\Gamma = 2$.
We show the bin-by-bin integrated energy-flux 95\% confidence level upper limits for each dSph candidate in \figref{sed}.
The Poisson likelihoods from each bin were combined to form global spectral likelihoods for different DM annihilation channels and masses.

We tested for excess gamma-ray emission consistent with two representative dark matter annihilation channels (\ie, \bbbar and \tautau) and a range of particle masses from 2 \GeV to 10 \TeV (when kinematically allowed).
No significant excess gamma-ray emission was observed from any of the DES dSph candidates for any of the DM masses or channels tested. The data were found to be well described by the background model with no significant residuals observed. 
We calculated the test statistic (\TS) for signal detection by comparing the likelihood values both with and without the added dSph candidate template (see Equation 6 in~\citet{Ackermann:2015zua}). 

The most significant excess, $\TS = 6.8$, was for \DESJa and a DM particle with $\mDM=25\GeV$ annihilating into \tautau.\footnote{We note that the radio-continuum source PMN\,J0335$-$5406 is located $\roughly 0\fdg1$ from the center of \DESJa. It is not a cataloged blazar, but has radio and infrared spectral characteristics consistent with blazars detected by the LAT.}
To convert from \TS to a local p-value, we use the \TS distribution measured by performing our search for gamma-ray emission in 4000 random blank sky fields~\citep{Ackermann:2013yva,Ackermann:2015zua}.\footnote{Though our blank sky ROIs are not independent, the overlap is negligible since we are testing for a point source at the center of the ROI. We have verified with a Monte Carlo all-sky realization that the TS distribution from our blank-sky analysis follows the asymptotic expectation when the background model perfectly describes the data.} 
We find that $\TS = 6.8$ corresponds to a local significance of $2.4\sigma$ ($p = 0.01$). After applying a trials factor to account for our scan in mass and annihilation channel, we calculate a significance of $1.65\sigma$ ($p = 0.05$) for this target. The global significance when accounting for fitting eight target locations is $0.43\sigma$ ($p = 0.33$).

Following the procedure described in the supplemental material of \citet{Ackermann:2015zua}, we investigated the systematic uncertainties related to uncertainties in the diffuse emission model by refitting with eight alternative diffuse models~\citep{dePalma:2013pia}. We found that using the alternative diffuse models varied the calculated limits and TS values by $\lesssim 20\%$. 

\section{Estimating J-factors for the DES dSph Candidates}\label{sec:photoJ}

The DM content of the DES dSph candidates cannot be determined without spectroscopic observations of their member stars.
However, it is possible to \New{predict the upper limits on the DM annihilation cross section that would be obtained given such observations by making the assumption} that these candidates possess DM distributions similar to the known dSphs.
Our estimates for the astrophysical J-factors of these candidates are motivated by two established relationships.
First, the known dSphs have a common mass scale in their interiors, roughly $10^7$ M$_\odot$ within their central 300 pc~\citep{Strigari:2008ib}. 
This radius is representative of the half light radius for classical dSphs, but is outside the visible stellar distribution of several ultra-faint satellites.
More generally, the half-light radius of a dSph and the mass within the half-light radius have been found to obey a simple scaling relation, assuming that the velocity dispersions are nearly constant in radius and the anisotropy of the stars is not strongly radially dependent~\citep{Walker:2009zp,Wolf:2009tu}.

\begin{figure}
  \begin{center}
  \includegraphics[width=\columnwidth]{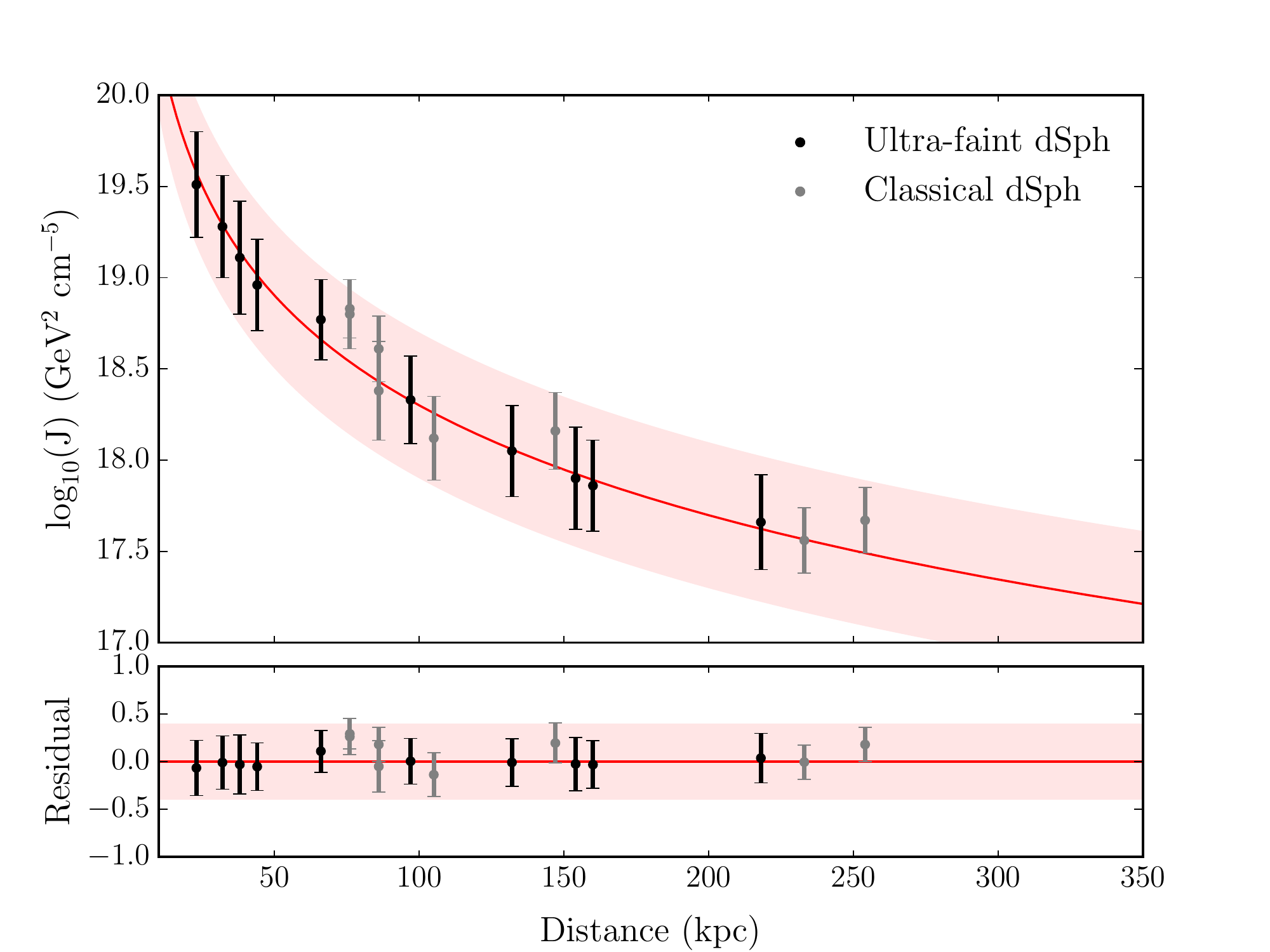}
  \caption{\label{fig:j_factor_scaling} \Jfactor distance scaling. Black points are from Table~1 in \citet{Ackermann:2013yva}. The red curve is our best-fit with an assumed inverse square distance relation (see text). The red band shows the $\pm0.4 \dex$ uncertainty that we adopt.}
\end{center}
\end{figure}

In the analysis that follows, we used the ten ultra-faint SDSS satellites with spectroscopically determined \Jfactors as a representative set of known dSphs.
Specifically, we take the \Jfactors calculated assuming an NFW profile integrated over a radius of 0\fdg5 for Bo\"{o}tes~I, Canes Venatici~I, Canes Venatici~II, Coma Berenices, Hercules, Leo~IV, Segue~1, Ursa Major~I, Ursa Major~II, and Willman~1 (see Table~1 in \citealt{Ackermann:2013yva}).
\figref{j_factor_scaling} shows the relation between the heliocentric distances and \Jfactors of ultra-faint and classical dSphs.
As expected from their similar interior DM masses, the \Jfactors of the known dSphs scale approximately as the inverse square of the distance.
The best-fit normalization is $\log_{10}({\rm J}) = 18.3\pm0.1$ at $d=100\kpc$.
We obtain a similar best-fit value, $\log_{10}({\rm J}) = 18.1\pm0.1$ at $d=100\kpc$, using the \Jfactors derived by~\citet{Geringer-Sameth:2014yza}, who assumed a generalized NFW profile and omitted Willman~1.\footnote{When using the values derived by~\citet{Geringer-Sameth:2014yza} and including Segue 2, we find a best-fit normalization of $\log_{10}({\rm J}) = 18.0\pm0.1$ at $d=100\kpc$.}
We note that the limited scatter in \figref{j_factor_scaling} is primarily due to the known dSphs residing in similar DM halos~\citep{Ackermann:2013yva}.
Under the assumption that the new DES dSph candidates belong to the same population, we estimated their \Jfactors based on the distances derived from the DES photometry.
\tabref{DESdwarfs} gives the estimated \Jfactors integrated over a solid-angle of $\Delta\Omega \sim 2.4 \times 10^{-4} \sr$ using our simple, empirical relation.

Several caveats should be noted.
None of the DES candidates have been confirmed to be gravitationally bound.
It is  possible that some have stellar populations characteristic of galaxies but lack substantial DM content, as is the case for Segue 2 \citep{Kirby:2013isa}, or have complicated kinematics that are difficult to interpret~\citep{Willman:2010gy}. 
Further, some of the M31 dSphs have been found to deviate from these relations, though it is possible that these deviations are due to tidal disruption~\citep{Collins:2013eek}.
Kinematic measurements of the member stars are needed to unambiguously resolve these questions.

\begin{figure*}
  \includegraphics[width=0.48\textwidth]{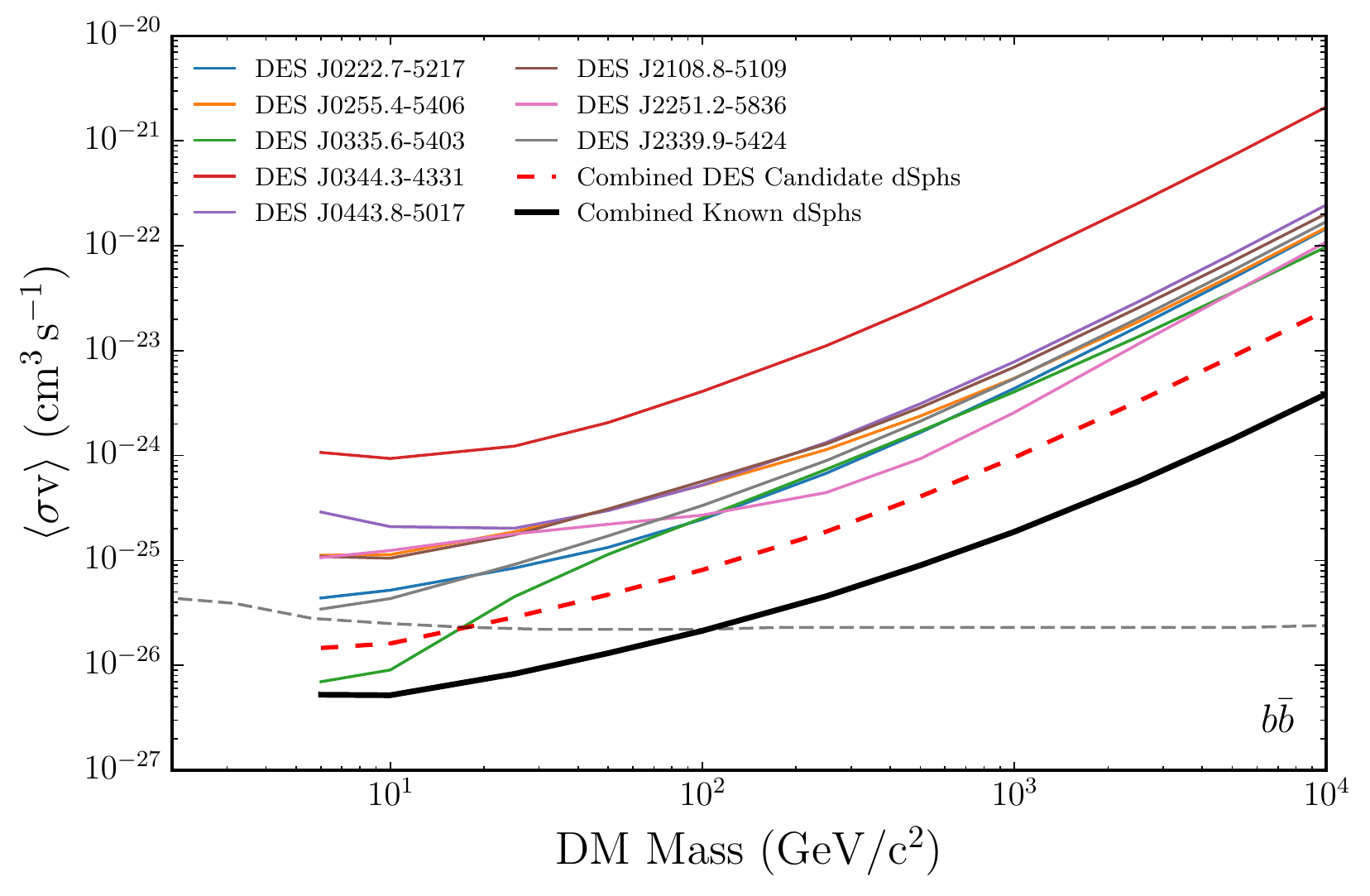} 
  \includegraphics[width=0.48\textwidth]{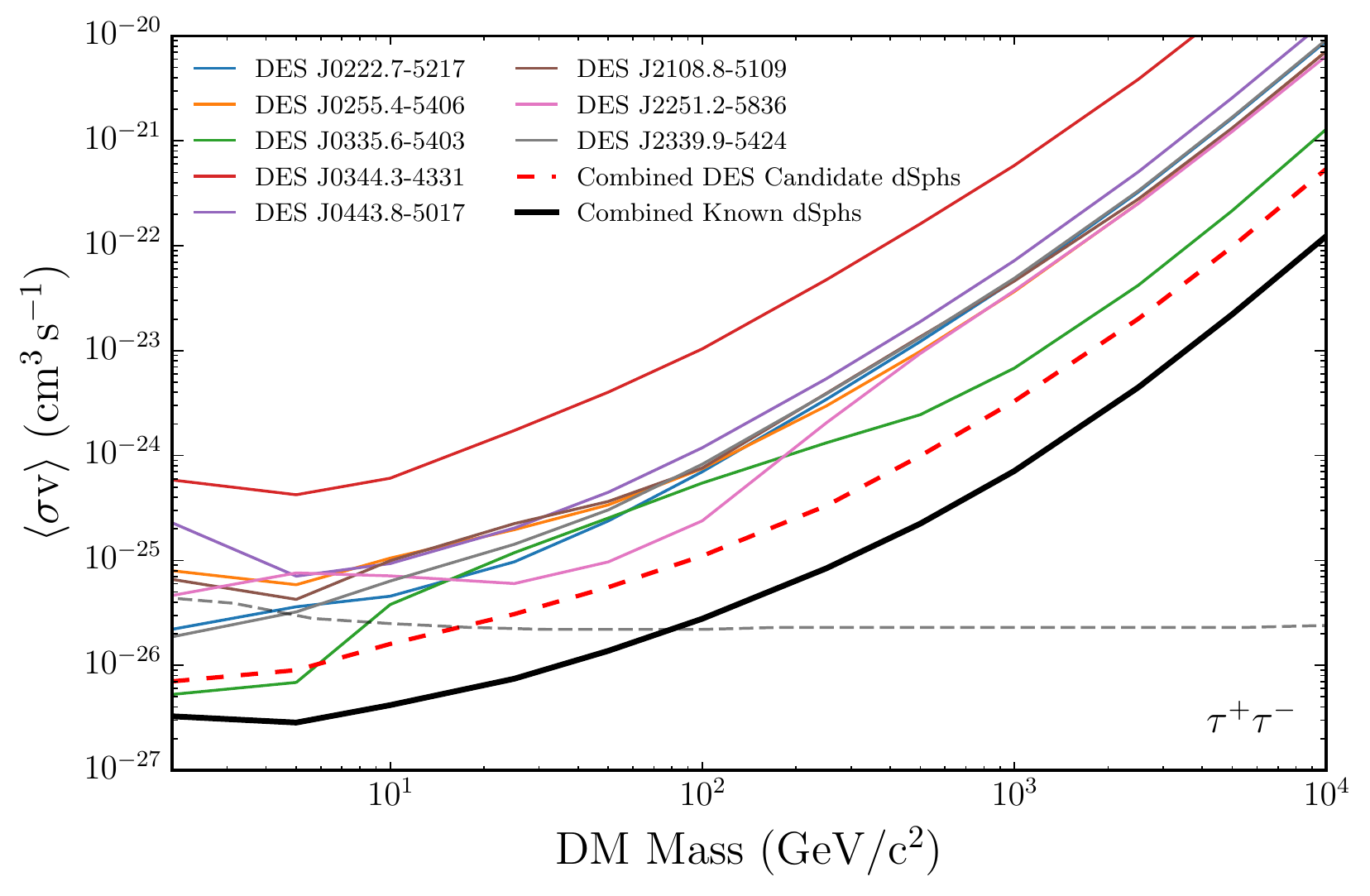} 
  \caption{\label{fig:DMlimits}
  Upper limits on the velocity-averaged DM annihilation cross section
  at 95\% confidence level for DM annihilation to \bbbar (left)
  and \tautau (right) derived using distance-estimated
  J-factors. Individual limits for each DES candidate dSph, as
  well as the combined limits (dashed red line) from the \nobjs new
  candidates are shown. Here we assume that each candidate is a dSph
  and \New{that future kinematic analyses will confirm the J-factors estimated based on photometric data (see text).} 
  For reference, we show the current best limits derived from a
  joint analysis of fifteen previously known dSphs with kinematically
  constrained \Jfactors (black curve) \citep{Ackermann:2015zua}. The
  dashed gray curve shows the thermal relic cross section derived
  by \citet{Steigman:2012nb}. }
\end{figure*}

Using the \Jfactor estimates presented in \tabref{DESdwarfs}, we followed the likelihood procedure detailed in \citet{Ackermann:2015zua} to obtain limits on DM annihilation from these \nobjs candidates shown in \figref{DMlimits}. 

We assumed a symmetric logarithmic uncertainty on the \Jfactor of $\pm 0.4 \dex$ for each DES candidate.
This value is representative of the uncertainties from ultra-faint dSphs \citep{Ackermann:2011wa,Geringer-Sameth:2014yza} and is somewhat larger than the uncertainties derived in \citet{Martinez:2013ioa}. 
The $\pm 0.4 \dex$ uncertainty is intended to represent the expected measurement uncertainty on the \Jfactors of the DES candidates after kinematic follow up.
The corresponding uncertainty band is illustrated in \figref{j_factor_scaling}.
We apply the same methodology as \citet{Ackermann:2015zua} to account for the \Jfactor uncertainty on each DES candidate by modeling it as a log normal distribution with ${\rm J}_{{\rm obs},i}$ equal to the values in \tabref{DESdwarfs}, and $\sigma_i = 0.4$ dex (see Equation 3 of \citet{Ackermann:2015zua}).

We derived individual and combined limits on the DM annihilation cross section for DM annihilation via the \bbbar and \tautau channels, under the assumption that each DES candidate is a dSph and has the \Jfactor listed in \tabref{DESdwarfs}.
We note that when using a \Jfactor uncertainty of $\pm 0.6 \dex$ instead of $\pm 0.4 \dex$, the individual dwarf candidate limits worsen by a factor of $\sim1.6$, while the combined limits worsen by 15--20\%.
We stress that the distance-estimated limits may differ substantially as spectroscopic data become available to more robustly constrain the DM content of the DES candidates.
However, once measured \Jfactors are obtained, the observed limits from each candidate will scale linearly with the measured \Jfactor relative to our estimates. 
Given the current uncertainty regarding the nature of the dSph candidates, we do not combine limits with those from previously known dSphs~(\ie, \citet{Ackermann:2015zua}).

\section{Discussion and Conclusions}\label{sec:sum}

The discovery of \nobjs dSph candidates in the first year of DES observations sets an optimistic tone for future dSph detections from DES and other optical surveys. 
\DESJa, at a distance of $\sim32$ kpc, is a particularly interesting candidate in this context, and should be considered a high-priority target for spectroscopic follow up.
The location of any newly discovered dSph, including the candidates investigated in this work, will have already been regularly observed since the beginning of the \Fermi mission.
No significant gamma-ray excess was found coincident with any of the \nobjs new DES dSph candidates considered here.
If \New{kinematic analyses find the dSph candidates to have \Jfactors similar to our estimates}, they constrain the annihilation cross section to lie below the thermal relic cross section for DM particles with masses $\lesssim 20 \GeV$ annihilating via the \bbbar or \tautau channels.

The population of nearby DM-dominated dSphs represents an independent set of targets to test possible signals of DM annihilation in other regions such as the Galactic center \citep[\eg,][]{Gordon:2013vta,Abazajian:2014fta,Daylan:2014rsa,Calore:2014xka}.
Though the expected DM signals of individual dSphs are smaller than that of the Galactic center, a joint-likelihood analysis of many dSphs can probe the DM annihilation cross section at a similar level of sensitivity.
The incorporation of new dSphs in indirect searches for DM with the LAT will further enhance the sensitivity of this method.

Independent analyses of \DESJa have been performed by \citet{Geringer-Sameth:2015lua} and \citet{Hooper:2015ula}. While the analysis details differ (\eg, the data set, the search technique, statistical methodology, and the calculation of the trials factor), each analysis finds the largest TS value in the direction of \DESJa. The p-values derived in \citeauthor{Geringer-Sameth:2015lua} and \citeauthor{Hooper:2015ula} are \New{smaller} than those found in this work.
One key difference is that~\citeauthor{Geringer-Sameth:2015lua} and \citeauthor{Hooper:2015ula} use the publicly available \irf{Pass 7 Reprocessed} data, while the analysis presented here uses the soon-to-be-released \irf{Pass 8} data, which improves the point-source sensitivity by 30\%--50\% in the relevant energy range.

\section*{Acknowledgments}

The \textit{Fermi}-LAT Collaboration acknowledges support for LAT development, operation and data analysis from NASA and DOE (United States), CEA/Irfu and IN2P3/CNRS (France), ASI and INFN (Italy), MEXT, KEK, and JAXA (Japan), and the K.A.~Wallenberg Foundation, the Swedish Research Council and the National Space Board (Sweden). Science analysis support in the operations phase from INAF (Italy) and CNES (France) is also gratefully acknowledged.

Funding for the DES Projects has been provided by the U.S. Department of Energy, the U.S. National Science Foundation, the Ministry of Science and Education of Spain, 
the Science and Technology Facilities Council of the United Kingdom, the Higher Education Funding Council for England, the National Center for Supercomputing 
Applications at the University of Illinois at Urbana-Champaign, the Kavli Institute of Cosmological Physics at the University of Chicago, 
the Center for Cosmology and Astro-Particle Physics at the Ohio State University,
the Mitchell Institute for Fundamental Physics and Astronomy at Texas A\&M University, Financiadora de Estudos e Projetos, 
Funda{\c c}{\~a}o Carlos Chagas Filho de Amparo {\`a} Pesquisa do Estado do Rio de Janeiro, Conselho Nacional de Desenvolvimento Cient{\'i}fico e Tecnol{\'o}gico and 
the Minist{\'e}rio da Ci{\^e}ncia, Tecnologia e Inova{\c c}{\~a}o, the Deutsche Forschungsgemeinschaft and the Collaborating Institutions in the Dark Energy Survey. 
The DES data management system is supported by the National Science Foundation under Grant Number AST-1138766.
The DES participants from Spanish institutions are partially supported by MINECO under grants AYA2012-39559, ESP2013-48274, FPA2013-47986, and Centro de Excelencia Severo Ochoa SEV-2012-0234, 
some of which include ERDF funds from the European Union.

The Collaborating Institutions are Argonne National Laboratory, the University of California at Santa Cruz, the University of Cambridge, Centro de Investigaciones En{\'e}rgeticas, 
Medioambientales y Tecnol{\'o}gicas-Madrid, the University of Chicago, University College London, the DES-Brazil Consortium, the University of Edinburgh, 
the Eidgen{\"o}ssische Technische Hochschule (ETH) Z{\"u}rich, 
Fermi National Accelerator Laboratory, the University of Illinois at Urbana-Champaign, the Institut de Ci{\`e}ncies de l'Espai (IEEC/CSIC), 
the Institut de F{\'i}sica d'Altes Energies, Lawrence Berkeley National Laboratory, the Ludwig-Maximilians Universit{\"a}t M{\"u}nchen and the associated Excellence Cluster Universe, 
the University of Michigan, the National Optical Astronomy Observatory, the University of Nottingham, The Ohio State University, the University of Pennsylvania, the University of Portsmouth, 
SLAC National Accelerator Laboratory, Stanford University, the University of Sussex, and Texas A\&M University.

ACR acknowledges financial support provided by the PAPDRJ CAPES/FAPERJ Fellowship.
AAP was supported by DOE grant DE-AC02-98CH10886 and by JPL, run by Caltech under a contract for NASA.
This research has made use of the NASA/IPAC Extragalactic Database (NED) which is operated by the Jet Propulsion Laboratory, California Institute of Technology, under contract with the National Aeronautics and Space Administration.
We would like to thank the anonomous referee for many helpful comments.

Facilities: Blanco, Fermi-LAT

\ifdefined\arxiv
  \bibliographystyle{apsrev4-1}
\else
  \bibliographystyle{apj}
\fi
\bibliography{bib}

\end{document}

%% file: commands.tex
\providecommand\fdg{\mbox{$.\!\!^\circ$}}

\newcommand{\nobjs}{{eight}\xspace}

\newcommand{\DESJa}{{DES\,J0335.6\allowbreak$-$5403}\xspace}
\newcommand{\DESJb}{{DES\,J0344.3\allowbreak$-$4331}\xspace}
\newcommand{\DESJc}{{DES\,J0255.4\allowbreak$-$5406}\xspace}
\newcommand{\DESJd}{{DES\,J0443.8\allowbreak$-$5017}\xspace}
\newcommand{\DESJe}{{DES\,J2251.2\allowbreak$-$5836}\xspace}
\newcommand{\DESJf}{{DES\,J0222.7\allowbreak$-$5217}\xspace}
\newcommand{\DESJg}{{DES\,J2339.9\allowbreak$-$5424}\xspace}
\newcommand{\DESJh}{{DES\,J2108.8\allowbreak$-$5109}\xspace}

\newcommand{\irf}[1]{\texttt{#1}}
\newcommand{\stools}{\emph{ScienceTools}}

\newcommand{\ie}{i.e.\xspace}
\newcommand{\eg}{e.g.\xspace}
\newcommand{\New}[1]{{#1}}

\mathchardef\mhyphen="2D

\newcommand{\vect}[1]{\boldsymbol{#1}}
\newcommand{\roughly}{\ensuremath{ {\sim}\,} }

\newcommand{\unit}[1]{\ensuremath{\mathrm{\,#1}}\xspace}
\newcommand{\dex}{\unit{dex}}
\newcommand{\MeV}{\unit{MeV}}
\newcommand{\GeV}{\unit{GeV}}
\newcommand{\TeV}{\unit{TeV}}
\newcommand{\degree}{\ensuremath{{}^{\circ}}\xspace}

\newcommand{\cm}{\unit{cm}}

\newcommand{\kpc}{\unit{kpc}}
\newcommand{\second}{\unit{s}}

\newcommand{\sr}{\unit{sr}}

\newcommand{\Gyr}{\unit{Gyr}\xspace}

\newcommand{\tabref}[1]{Table~\ref{tab:#1}}

\newcommand{\figref}[1]{Figure~\ref{fig:#1}}

\newcommand{\Fermi}{\textit{Fermi}\xspace}

\newcommand{\TS}{\ensuremath{\mathrm{TS}}\xspace}

\newcommand{\code}[1]{\lstinline!#1!\xspace}

\newcommand{\DM}{\ensuremath{\mathrm{DM}}}
\newcommand{\mDM}{\ensuremath{m_\DM}\xspace}

\newcommand{\sigmav}{\ensuremath{\langle \sigma v \rangle}\xspace}

\newcommand{\bbbar}{\ensuremath{b \bar b}\xspace}

\newcommand{\tautau}{\ensuremath{\tau^{+}\tau^{-}}\xspace}
\newcommand{\relic}{\ensuremath{2.2\times10^{-26}\cm^{3}\second^{-1}}\xspace}

\newcommand{\Jfactor}{\ensuremath{\mathrm{J\mhyphen factor}}\xspace}
\newcommand{\Jfactors}{\ensuremath{\mathrm{J\mhyphen factors}}\xspace}

\newlength{\twocolfigwidth}
\newlength{\onecolfigwidth}
\newlength{\onecolfigwidths}

%% file: authors_revtex.tex

\author{A.~Drlica-Wagner}
\email{kadrlica@fnal.gov}
\affiliation{Member of Fermi-LAT and DES collaborations.}
\affiliation{Fermi National Accelerator Laboratory, P. O. Box 500, Batavia, IL 60510, USA}
\author{A.~Albert}
\email{aalbert@slac.stanford.edu}
\affiliation{W. W. Hansen Experimental Physics Laboratory, Kavli Institute for Particle Astrophysics and Cosmology, Department of Physics and SLAC National Accelerator Laboratory, Stanford University, Stanford, CA 94305, USA}
\author{K.~Bechtol}
\email{bechtol@kicp.uchicago.edu}
\affiliation{Member of Fermi-LAT and DES collaborations.}
\affiliation{Kavli Institute for Cosmological Physics, University of Chicago, Chicago, IL 60637, USA}
\author{M.~Wood}
\email{mdwood@slac.stanford.edu}
\affiliation{W. W. Hansen Experimental Physics Laboratory, Kavli Institute for Particle Astrophysics and Cosmology, Department of Physics and SLAC National Accelerator Laboratory, Stanford University, Stanford, CA 94305, USA}
\author{L.~Strigari}
\email{strigari@physics.tamu.edu}
\affiliation{George P. and Cynthia Woods Mitchell Institute for Fundamental Physics and Astronomy, and Department of Physics and Astronomy, Texas A\&M University, College Station, TX 77843,  USA}
\author{M.~S\'anchez-Conde}
\affiliation{The Oskar Klein Centre for Cosmoparticle Physics, AlbaNova, SE-106 91 Stockholm, Sweden}
\affiliation{Department of Physics, Stockholm University, AlbaNova, SE-106 91 Stockholm, Sweden}
\author{L.~Baldini}
\affiliation{Universit\`a di Pisa and Istituto Nazionale di Fisica Nucleare, Sezione di Pisa I-56127 Pisa, Italy}
\author{R.~Essig}
\affiliation{C.N. Yang Institute for Theoretical Physics, State University of New York, Stony Brook, NY 11794-3840, U.S.A., USA}
\author{J.~Cohen-Tanugi}
\affiliation{Laboratoire Univers et Particules de Montpellier, Universit\'e Montpellier 2, CNRS/IN2P3, Montpellier, France}
\author{B.~Anderson}
\affiliation{Royal Swedish Academy of Sciences Research Fellow, funded by a grant from the K. A. Wallenberg Foundation}
\author{R.~Bellazzini}
\affiliation{Istituto Nazionale di Fisica Nucleare, Sezione di Pisa, I-56127 Pisa, Italy}
\author{E.~D.~Bloom}
\affiliation{W. W. Hansen Experimental Physics Laboratory, Kavli Institute for Particle Astrophysics and Cosmology, Department of Physics and SLAC National Accelerator Laboratory, Stanford University, Stanford, CA 94305, USA}
\author{R.~Caputo}
\affiliation{Santa Cruz Institute for Particle Physics, Department of Physics and Department of Astronomy and Astrophysics, University of California at Santa Cruz, Santa Cruz, CA 95064, USA}
\author{C.~Cecchi}
\affiliation{Istituto Nazionale di Fisica Nucleare, Sezione di Perugia, I-06123 Perugia, Italy}
\affiliation{Dipartimento di Fisica, Universit\`a degli Studi di Perugia, I-06123 Perugia, Italy}
\author{E.~Charles}
\affiliation{W. W. Hansen Experimental Physics Laboratory, Kavli Institute for Particle Astrophysics and Cosmology, Department of Physics and SLAC National Accelerator Laboratory, Stanford University, Stanford, CA 94305, USA}
\author{J.~Chiang}
\affiliation{W. W. Hansen Experimental Physics Laboratory, Kavli Institute for Particle Astrophysics and Cosmology, Department of Physics and SLAC National Accelerator Laboratory, Stanford University, Stanford, CA 94305, USA}
\author{A.~de~Angelis}
\affiliation{Dipartimento di Fisica, Universit\`a di Udine and Istituto Nazionale di Fisica Nucleare, Sezione di Trieste, Gruppo Collegato di Udine, I-33100 Udine}
\author{S.~Funk}
\affiliation{W. W. Hansen Experimental Physics Laboratory, Kavli Institute for Particle Astrophysics and Cosmology, Department of Physics and SLAC National Accelerator Laboratory, Stanford University, Stanford, CA 94305, USA}
\author{P.~Fusco}
\affiliation{Dipartimento di Fisica ``M. Merlin" dell'Universit\`a e del Politecnico di Bari, I-70126 Bari, Italy}
\affiliation{Istituto Nazionale di Fisica Nucleare, Sezione di Bari, 70126 Bari, Italy}
\author{F.~Gargano}
\affiliation{Istituto Nazionale di Fisica Nucleare, Sezione di Bari, 70126 Bari, Italy}
\author{N.~Giglietto}
\affiliation{Dipartimento di Fisica ``M. Merlin" dell'Universit\`a e del Politecnico di Bari, I-70126 Bari, Italy}
\affiliation{Istituto Nazionale di Fisica Nucleare, Sezione di Bari, 70126 Bari, Italy}
\author{F.~Giordano}
\affiliation{Dipartimento di Fisica ``M. Merlin" dell'Universit\`a e del Politecnico di Bari, I-70126 Bari, Italy}
\affiliation{Istituto Nazionale di Fisica Nucleare, Sezione di Bari, 70126 Bari, Italy}
\author{S.~Guiriec}
\affiliation{NASA Goddard Space Flight Center, Greenbelt, MD 20771, USA}
\affiliation{NASA Postdoctoral Program Fellow, USA}
\author{M.~Gustafsson}
\affiliation{Georg-August University G\"ottingen, Institute for theoretical Physics - Faculty of Physics, Friedrich-Hund-Platz 1, D-37077 G\"ottingen, Germany}
\author{M.~Kuss}
\affiliation{Istituto Nazionale di Fisica Nucleare, Sezione di Pisa, I-56127 Pisa, Italy}
\author{F.~Loparco}
\affiliation{Dipartimento di Fisica ``M. Merlin" dell'Universit\`a e del Politecnico di Bari, I-70126 Bari, Italy}
\affiliation{Istituto Nazionale di Fisica Nucleare, Sezione di Bari, 70126 Bari, Italy}
\author{P.~Lubrano}
\affiliation{Istituto Nazionale di Fisica Nucleare, Sezione di Perugia, I-06123 Perugia, Italy}
\affiliation{Dipartimento di Fisica, Universit\`a degli Studi di Perugia, I-06123 Perugia, Italy}
\author{N.~Mirabal}
\affiliation{NASA Goddard Space Flight Center, Greenbelt, MD 20771, USA}
\affiliation{NASA Postdoctoral Program Fellow, USA}
\author{T.~Mizuno}
\affiliation{Hiroshima Astrophysical Science Center, Hiroshima University, Higashi-Hiroshima, Hiroshima 739-8526, Japan}
\author{A.~Morselli}
\affiliation{Istituto Nazionale di Fisica Nucleare, Sezione di Roma ``Tor Vergata", I-00133 Roma, Italy}
\author{T.~Ohsugi}
\affiliation{Hiroshima Astrophysical Science Center, Hiroshima University, Higashi-Hiroshima, Hiroshima 739-8526, Japan}
\author{E.~Orlando}
\affiliation{W. W. Hansen Experimental Physics Laboratory, Kavli Institute for Particle Astrophysics and Cosmology, Department of Physics and SLAC National Accelerator Laboratory, Stanford University, Stanford, CA 94305, USA}
\author{M.~Persic}
\affiliation{Istituto Nazionale di Fisica Nucleare, Sezione di Trieste, I-34127 Trieste, Italy}
\affiliation{Osservatorio Astronomico di Trieste, Istituto Nazionale di Astrofisica, I-34143 Trieste, Italy}
\author{S.~Rain\`o}
\affiliation{Dipartimento di Fisica ``M. Merlin" dell'Universit\`a e del Politecnico di Bari, I-70126 Bari, Italy}
\affiliation{Istituto Nazionale di Fisica Nucleare, Sezione di Bari, 70126 Bari, Italy}
\author{N.~Sehgal}
\affiliation{Physics and Astronomy Department, Stony Brook University, Stony Brook, NY 11794, USA}
\author{F.~Spada}
\affiliation{Istituto Nazionale di Fisica Nucleare, Sezione di Pisa, I-56127 Pisa, Italy}
\author{D.~J.~Suson}
\affiliation{Department of Chemistry and Physics, Purdue University Calumet, Hammond, IN 46323-2094, USA}
\author{G.~Zaharijas}
\affiliation{Istituto Nazionale di Fisica Nucleare, Sezione di Trieste, and Universit\`a di Trieste, I-34127 Trieste, Italy}
\affiliation{Laboratory for Astroparticle Physics, University of Nova Gorica, Vipavska 13, SI-5000 Nova Gorica, Slovenia}
\author{S.~Zimmer}
\affiliation{Department of Physics, Stockholm University, AlbaNova, SE-106 91 Stockholm, Sweden}
\affiliation{The Oskar Klein Centre for Cosmoparticle Physics, AlbaNova, SE-106 91 Stockholm, Sweden}

\collaboration{The Fermi-LAT Collaboration}


\author{T.~Abbott}
\affiliation{Cerro Tololo Inter-American Observatory, National Optical Astronomy Observatory, Casilla 603, La Serena, Chile}
\author{S.~Allam}
\affiliation{Fermi National Accelerator Laboratory, P. O. Box 500, Batavia, IL 60510, USA}
\affiliation{Space Telescope Science Institute, 3700 San Martin Drive, Baltimore, MD  21218, USA}
\author{E.~Balbinot}
\affiliation{Department of Physics, University of Surrey, Guildford GU2 7XH, UK}
\affiliation{Laborat\'orio Interinstitucional de e-Astronomia - LIneA, Rua Gal. Jos\'e Cristino 77, Rio de Janeiro, RJ - 20921-400, Brazil}
\author{A.~H.~Bauer}
\affiliation{Institut de Ci\`encies de l'Espai, IEEC-CSIC, Campus UAB, Facultat de Ci\`encies, Torre C5 par-2, 08193 Bellaterra, Barcelona, Spain}
\author{A.~Benoit-L{\'e}vy}
\affiliation{Department of Physics \& Astronomy, University College London, Gower Street, London, WC1E 6BT, UK}
\author{R.~A.~Bernstein}
\affiliation{Carnegie Observatories, 813 Santa Barbara St., Pasadena, CA 91101, USA}
\author{G.~M.~Bernstein}
\affiliation{Department of Physics and Astronomy, University of Pennsylvania, Philadelphia, PA 19104, USA}
\author{E.~Bertin}
\affiliation{Institut d'Astrophysique de Paris, Univ. Pierre et Marie Curie \& CNRS UMR7095, F-75014 Paris, France}
\author{D.~Brooks}
\affiliation{Department of Physics \& Astronomy, University College London, Gower Street, London, WC1E 6BT, UK}
\author{E.~Buckley-Geer}
\affiliation{Fermi National Accelerator Laboratory, P. O. Box 500, Batavia, IL 60510, USA}
\author{D.~L.~Burke}
\affiliation{SLAC National Accelerator Laboratory, Menlo Park, CA 94025, USA}
\author{A.~Carnero Rosell}
\affiliation{Observat\'orio Nacional, Rua Gal. Jos\'e Cristino 77, Rio de Janeiro, RJ - 20921-400, Brazil}
\affiliation{Laborat\'orio Interinstitucional de e-Astronomia - LIneA, Rua Gal. Jos\'e Cristino 77, Rio de Janeiro, RJ - 20921-400, Brazil}
\author{F.~J.~Castander}
\affiliation{Institut de Ci\`encies de l'Espai, IEEC-CSIC, Campus UAB, Facultat de Ci\`encies, Torre C5 par-2, 08193 Bellaterra, Barcelona, Spain}
\author{R.~Covarrubias}
\affiliation{National Center for Supercomputing Applications, 1205 West Clark St., Urbana, IL 61801, USA}
\author{C.~B.~D'Andrea}
\affiliation{Institute of Cosmology \& Gravitation, University of Portsmouth, Portsmouth, PO1 3FX, UK}
\author{L.~N.~da~Costa}
\affiliation{Laborat\'orio Interinstitucional de e-Astronomia - LIneA, Rua Gal. Jos\'e Cristino 77, Rio de Janeiro, RJ - 20921-400, Brazil}
\affiliation{Observat\'orio Nacional, Rua Gal. Jos\'e Cristino 77, Rio de Janeiro, RJ - 20921-400, Brazil}
\author{D.~L.~DePoy}
\affiliation{George P. and Cynthia Woods Mitchell Institute for Fundamental Physics and Astronomy, and Department of Physics and Astronomy, Texas A\&M University, College Station, TX 77843,  USA}
\author{S.~Desai}
\affiliation{Department of Physics, Ludwig-Maximilians-Universitat, Scheinerstr.\ 1, 81679 Munich, Germany}
\affiliation{Excellence Cluster Universe, Boltzmannstr.\ 2, 85748 Garching, Germany}
\author{H.~T.~Diehl}
\affiliation{Fermi National Accelerator Laboratory, P. O. Box 500, Batavia, IL 60510, USA}
\author{C.~E Cunha}
\affiliation{Kavli Institute for Particle Astrophysics \& Cosmology, P. O. Box 2450, Stanford University, Stanford, CA 94305, USA}
\author{T.~F.~Eifler}
\affiliation{Department of Physics and Astronomy, University of Pennsylvania, Philadelphia, PA 19104, USA}
\affiliation{Jet Propulsion Laboratory, California Institute of Technology, 4800 Oak Grove Dr., Pasadena, CA 91109, USA}
\author{J.~Estrada}
\affiliation{Fermi National Accelerator Laboratory, P. O. Box 500, Batavia, IL 60510, USA}
\author{A.~E.~Evrard}
\affiliation{Department of Physics, University of Michigan, Ann Arbor, MI 48109, USA}
\author{A.~Fausti Neto}
\affiliation{Laborat\'orio Interinstitucional de e-Astronomia - LIneA, Rua Gal. Jos\'e Cristino 77, Rio de Janeiro, RJ - 20921-400, Brazil}
\author{E.~Fernandez}
\affiliation{Institut de F\'{\i}sica d'Altes Energies, Universitat Aut\`onoma de Barcelona, E-08193 Bellaterra, Barcelona, Spain}
\author{D.~A.~Finley}
\affiliation{Fermi National Accelerator Laboratory, P. O. Box 500, Batavia, IL 60510, USA}
\author{B.~Flaugher}
\affiliation{Fermi National Accelerator Laboratory, P. O. Box 500, Batavia, IL 60510, USA}
\author{J.~Frieman}
\affiliation{Fermi National Accelerator Laboratory, P. O. Box 500, Batavia, IL 60510, USA}
\affiliation{Kavli Institute for Cosmological Physics, University of Chicago, Chicago, IL 60637, USA}
\author{E.~Gaztanaga}
\affiliation{Institut de Ci\`encies de l'Espai, IEEC-CSIC, Campus UAB, Facultat de Ci\`encies, Torre C5 par-2, 08193 Bellaterra, Barcelona, Spain}
\author{D.~Gerdes}
\affiliation{Department of Physics, University of Michigan, Ann Arbor, MI 48109, USA}
\author{D.~Gruen}
\affiliation{Max Planck Institute for Extraterrestrial Physics, Giessenbachstrasse, 85748 Garching, Germany}
\affiliation{University Observatory Munich, Scheinerstrasse 1, 81679 Munich, Germany}
\author{R.~A.~Gruendl}
\affiliation{Department of Astronomy, University of Illinois,1002 W. Green Street, Urbana, IL 61801, USA}
\affiliation{National Center for Supercomputing Applications, 1205 West Clark St., Urbana, IL 61801, USA}
\author{G.~Gutierrez}
\affiliation{Fermi National Accelerator Laboratory, P. O. Box 500, Batavia, IL 60510, USA}
\author{K.~Honscheid}
\affiliation{Department of Physics, The Ohio State University, Columbus, OH 43210, USA}
\affiliation{Center for Cosmology and Astro-Particle Physics, The Ohio State University, Columbus, OH 43210, USA}
\author{B.~Jain}
\affiliation{Department of Physics and Astronomy, University of Pennsylvania, Philadelphia, PA 19104, USA}
\author{D.~James}
\affiliation{Cerro Tololo Inter-American Observatory, National Optical Astronomy Observatory, Casilla 603, La Serena, Chile}
\author{T.~Jeltema}
\affiliation{Santa Cruz Institute for Particle Physics, Department of Physics and Department of Astronomy and Astrophysics, University of California at Santa Cruz, Santa Cruz, CA 95064, USA}
\author{S.~Kent}
\affiliation{Fermi National Accelerator Laboratory, P. O. Box 500, Batavia, IL 60510, USA}
\author{R.~Kron}
\affiliation{Kavli Institute for Cosmological Physics, University of Chicago, Chicago, IL 60637, USA}
\author{K.~Kuehn}
\affiliation{Australian Astronomical Observatory, North Ryde, NSW 2113, Australia and Argonne National Laboratory, 9700 S. Cass Avenue, Lemont IL 60639 USA}
\author{N.~Kuropatkin}
\affiliation{Fermi National Accelerator Laboratory, P. O. Box 500, Batavia, IL 60510, USA}
\author{O.~Lahav}
\affiliation{Department of Physics \& Astronomy, University College London, Gower Street, London, WC1E 6BT, UK}
\author{T.~S.~Li}
\affiliation{George P. and Cynthia Woods Mitchell Institute for Fundamental Physics and Astronomy, and Department of Physics and Astronomy, Texas A\&M University, College Station, TX 77843,  USA}
\author{E.~Luque}
\affiliation{Instituto de F\'\i sica, UFRGS, Caixa Postal 15051, Porto Alegre, RS - 91501-970, Brazil}
\affiliation{Laborat\'orio Interinstitucional de e-Astronomia - LIneA, Rua Gal. Jos\'e Cristino 77, Rio de Janeiro, RJ - 20921-400, Brazil}
\author{M.~A.~G.~Maia}
\affiliation{Laborat\'orio Interinstitucional de e-Astronomia - LIneA, Rua Gal. Jos\'e Cristino 77, Rio de Janeiro, RJ - 20921-400, Brazil}
\affiliation{Observat\'orio Nacional, Rua Gal. Jos\'e Cristino 77, Rio de Janeiro, RJ - 20921-400, Brazil}
\author{M.~Makler}
\affiliation{ICRA, Centro Brasileiro de Pesquisas F\'isicas, Rua Dr. Xavier Sigaud 150, CEP 22290-180, Rio de Janeiro, RJ, Brazil}
\author{M.~March}
\affiliation{Department of Physics and Astronomy, University of Pennsylvania, Philadelphia, PA 19104, USA}
\author{J.~Marshall}
\affiliation{George P. and Cynthia Woods Mitchell Institute for Fundamental Physics and Astronomy, and Department of Physics and Astronomy, Texas A\&M University, College Station, TX 77843,  USA}
\author{P.~Martini}
\affiliation{Center for Cosmology and Astro-Particle Physics, The Ohio State University, Columbus, OH 43210, USA}
\affiliation{Department of Astronomy, The Ohio State University, Columbus, OH 43210, USA}
\author{K.~W.~Merritt}
\affiliation{Fermi National Accelerator Laboratory, P. O. Box 500, Batavia, IL 60510, USA}
\author{C.~Miller}
\affiliation{Department of Physics, University of Michigan, Ann Arbor, MI 48109, USA}
\author{R.~Miquel}
\affiliation{Institut de F\'{\i}sica d'Altes Energies, Universitat Aut\`onoma de Barcelona, E-08193 Bellaterra, Barcelona, Spain}
\affiliation{Instituci\'o Catalana de Recerca i Estudis Avan\c{c}ats, E-08010 Barcelona (Spain)}
\author{J.~Mohr}
\affiliation{Department of Physics, Ludwig-Maximilians-Universitat, Scheinerstr.\ 1, 81679 Munich, Germany}
\affiliation{Excellence Cluster Universe, Boltzmannstr.\ 2, 85748 Garching, Germany}
\affiliation{Max Planck Institute for Extraterrestrial Physics, Giessenbachstrasse, 85748 Garching, Germany}
\author{E.~Neilsen}
\affiliation{Fermi National Accelerator Laboratory, P. O. Box 500, Batavia, IL 60510, USA}
\author{B.~Nord}
\affiliation{Fermi National Accelerator Laboratory, P. O. Box 500, Batavia, IL 60510, USA}
\author{R.~Ogando}
\affiliation{Observat\'orio Nacional, Rua Gal. Jos\'e Cristino 77, Rio de Janeiro, RJ - 20921-400, Brazil}
\affiliation{Laborat\'orio Interinstitucional de e-Astronomia - LIneA, Rua Gal. Jos\'e Cristino 77, Rio de Janeiro, RJ - 20921-400, Brazil}
\author{J.~Peoples}
\affiliation{Fermi National Accelerator Laboratory, P. O. Box 500, Batavia, IL 60510, USA}
\author{D.~Petravick}
\affiliation{National Center for Supercomputing Applications, 1205 West Clark St., Urbana, IL 61801, USA}
\author{A.~Pieres}
\affiliation{Instituto de F\'\i sica, UFRGS, Caixa Postal 15051, Porto Alegre, RS - 91501-970, Brazil}
\affiliation{Laborat\'orio Interinstitucional de e-Astronomia - LIneA, Rua Gal. Jos\'e Cristino 77, Rio de Janeiro, RJ - 20921-400, Brazil}
\author{A.~A.~Plazas}
\affiliation{Brookhaven National Laboratory, Bldg 510, Upton, NY 11973, USA}
\affiliation{Jet Propulsion Laboratory, California Institute of Technology, 4800 Oak Grove Dr., Pasadena, CA 91109, USA}
\author{A.~Queiroz}
\affiliation{Instituto de F\'\i sica, UFRGS, Caixa Postal 15051, Porto Alegre, RS - 91501-970, Brazil}
\affiliation{Laborat\'orio Interinstitucional de e-Astronomia - LIneA, Rua Gal. Jos\'e Cristino 77, Rio de Janeiro, RJ - 20921-400, Brazil}
\author{A.~K.~Romer}
\affiliation{Astronomy Centre, University of Sussex, Falmer, Brighton, BN1 9QH, UK}
\author{A.~Roodman}
\affiliation{Kavli Institute for Particle Astrophysics \& Cosmology, P. O. Box 2450, Stanford University, Stanford, CA 94305, USA}
\affiliation{SLAC National Accelerator Laboratory, Menlo Park, CA 94025, USA}
\author{E.~S.~Rykoff}
\affiliation{SLAC National Accelerator Laboratory, Menlo Park, CA 94025, USA}
\author{M.~Sako}
\affiliation{Department of Physics and Astronomy, University of Pennsylvania, Philadelphia, PA 19104, USA}
\author{E.~Sanchez}
\affiliation{Centro de Investigaciones Energ\'eticas, Medioambientales y Tecnol\'ogicas (CIEMAT), Madrid, Spain}
\author{B.~Santiago}
\affiliation{Instituto de F\'\i sica, UFRGS, Caixa Postal 15051, Porto Alegre, RS - 91501-970, Brazil}
\affiliation{Laborat\'orio Interinstitucional de e-Astronomia - LIneA, Rua Gal. Jos\'e Cristino 77, Rio de Janeiro, RJ - 20921-400, Brazil}
\author{V.~Scarpine}
\affiliation{Fermi National Accelerator Laboratory, P. O. Box 500, Batavia, IL 60510, USA}
\author{M.~Schubnell}
\affiliation{Department of Physics, University of Michigan, Ann Arbor, MI 48109, USA}
\author{I.~Sevilla}
\affiliation{Centro de Investigaciones Energ\'eticas, Medioambientales y Tecnol\'ogicas (CIEMAT), Madrid, Spain}
\affiliation{Department of Astronomy, University of Illinois,1002 W. Green Street, Urbana, IL 61801, USA}
\author{R.~C.~Smith}
\affiliation{Cerro Tololo Inter-American Observatory, National Optical Astronomy Observatory, Casilla 603, La Serena, Chile}
\author{M.~Soares-Santos}
\affiliation{Fermi National Accelerator Laboratory, P. O. Box 500, Batavia, IL 60510, USA}
\author{F.~Sobreira}
\affiliation{Fermi National Accelerator Laboratory, P. O. Box 500, Batavia, IL 60510, USA}
\affiliation{Laborat\'orio Interinstitucional de e-Astronomia - LIneA, Rua Gal. Jos\'e Cristino 77, Rio de Janeiro, RJ - 20921-400, Brazil}
\author{E.~Suchyta}
\affiliation{Center for Cosmology and Astro-Particle Physics, The Ohio State University, Columbus, OH 43210, USA}
\affiliation{Department of Physics, The Ohio State University, Columbus, OH 43210, USA}
\author{M.~E.~C.~Swanson}
\affiliation{National Center for Supercomputing Applications, 1205 West Clark St., Urbana, IL 61801, USA}
\author{G.~Tarle}
\affiliation{Department of Physics, University of Michigan, Ann Arbor, MI 48109, USA}
\author{J.~Thaler}
\affiliation{Department of Physics, University of Illinois, 1110 W. Green St., Urbana, IL 61801, USA}
\author{D.~Thomas}
\affiliation{Institute of Cosmology \& Gravitation, University of Portsmouth, Portsmouth, PO1 3FX, UK}
\author{D.~Tucker}
\affiliation{Fermi National Accelerator Laboratory, P. O. Box 500, Batavia, IL 60510, USA}
\author{A.~Walker}
\affiliation{Cerro Tololo Inter-American Observatory, National Optical Astronomy Observatory, Casilla 603, La Serena, Chile}
\author{R.~H.~Wechsler}
\affiliation{Department of Physics, Stanford University, 382 Via Pueblo Mall, Stanford, CA 94305, USA}
\affiliation{Kavli Institute for Particle Astrophysics \& Cosmology, P. O. Box 2450, Stanford University, Stanford, CA 94305, USA}
\affiliation{SLAC National Accelerator Laboratory, Menlo Park, CA 94025, USA}
\author{W.~Wester}
\affiliation{Fermi National Accelerator Laboratory, P. O. Box 500, Batavia, IL 60510, USA}
\author{P.~Williams}
\affiliation{Kavli Institute for Cosmological Physics, University of Chicago, Chicago, IL 60637, USA}
\author{B.~Yanny}
\affiliation{Fermi National Accelerator Laboratory, P. O. Box 500, Batavia, IL 60510, USA}
\author{J.~Zuntz}
\affiliation{Jodrell Bank Center for Astrophysics, School of Physics and Astronomy, University of Manchester, Oxford Road, Manchester, M13 9PL, UK}

\collaboration{The DES Collaboration}